\documentclass[twocolumn,aps,prl,floatfix,showpacs]{revtex4}
\usepackage{graphicx,graphics,color,epsfig}
\usepackage{amsmath}
\usepackage{amssymb}

\newcommand{\be}{\begin{equation}}
\newcommand{\ee}{\end{equation}}

\begin{document}

\title{
Efficient simulation  of relativistic fermions via vertex models
}

\author{Urs Wenger}

\affiliation{
Institute for Theoretical Physics, University of Bern, Sidlerstrasse 5,
CH--3012 Bern, Switzerland
}

\date{\today}

\begin{abstract}
  We have developed an efficient simulation algorithm for strongly interacting
  relativistic fermions in two-dimensional field theories based on a
  formulation as a loop gas. The loop models describing the dynamics of the
  fermions can be mapped to statistical vertex models and our proposal is in
  fact an efficient simulation algorithm for generic vertex models in
  arbitrary dimensions. The algorithm essentially eliminates critical slowing
  down by sampling two-point correlation functions and it allows simulations
  directly in the massless limit. Moreover, it generates loop configurations
  with fluctuating topological boundary conditions enabling to simulate
  fermions with arbitrary periodic or anti-periodic boundary conditions. As
  illustrative examples, the algorithm is applied to the Gross-Neveu model and
  to the Schwinger model in the strong coupling limit.
\end{abstract}

\pacs{02.70.-c, 05.50.+q, 11.10.Kk, 11.15.Ha}

\maketitle

Simulating strongly interacting fermions, like in Quantum Chromodynamics (QCD)
or in Nambu-Jona--Lasinio models, is considered to be rather difficult and
continues to be a challenge.  Contrary to common lore, this is not directly
due to the fact that the fermionic degrees of freedom are Grassmannian
variables, but rather due to the non-locality of the determinant which is
obtained upon integrating out the fermionic fields. Moreover, simulations of
fermions are usually hampered by critical slowing down towards the chiral
limit where the fermions become massless and the correlation length of the
fermionic two-point function diverges.  The established standard method to
perform such calculations on the lattice is to use the Hybrid Monte-Carlo
algorithm \cite{Duane:1987de} which deals with the non-locality of the
determinant by rewriting it as an integral over bosonic "pseudo-fermion"
fields. The algorithm then requires to deal with the inverse of the fermion
Dirac operator, however, the operator becomes ill-conditioned towards the
massless limit and the simulations slow down dramatically.  In this letter we
propose a novel approach which circumvents the above mentioned problems. It is
based on a (high-temperature) expansion of the fermion actions which
reformulates the fermionic systems as $q$-state vertex models, i.e.,
statistical closed loop models. In particular, the method is directly
applicable to the Gross-Neveu (GN) model and to the Schwinger model in the
strong coupling limit. These models can be shown to be equivalent to specific
vertex models \cite{Salmhofer:1991cc,Scharnhorst:1996sv,Scharnhorst:1996gj,
  Gattringer:1998cd} and our simulation method, based on a proposal by
Prokof'ev and Svistunov \cite{Prokof'ev:2001zz}, is effectively a very
efficient updating algorithm for generic vertex models (in arbitrary
dimensions).  In fact, the algorithm essentially eliminates critical slowing
down and is able to simulate the fermionic systems at the critical point and
directly in the massless limit.

We start with illustrating the reformulation in terms of closed loops in the
GN model. The model is most naturally formulated by employing Majorana
fermions \cite{Itzykson:1989sy,Wolff:2007ip}. Here we are using Wilson's
Euclidean lattice discretisation for which the action density of the model is
\be 
{\cal L}_\textrm{GN} = \frac{1}{2} \xi^T {\cal C} (\gamma_\mu
\tilde \partial_\mu - \frac{1}{2} \partial^* \partial + m) \xi - \frac{g^2}{4}
\left( \xi^T {\cal C} \xi \right)^2, 
\ee 
where $\xi$ is a real, two component Grassmann field describing a Majorana
fermion with mass $m$, ${\cal C} = -{\cal C}^T$ is the charge conjugation
matrix, and $\partial, \partial^*, \tilde
\partial$ denote the forward, backward and symmetric lattice derivative,
respectively. The Wilson term $\frac{1}{2} \partial^* \partial$, responsible
for removing the fermion doublers, explicitly breaks the discrete chiral
symmetry $\xi \rightarrow \gamma_5 \xi, \xi^T {\cal C} \rightarrow \xi^T {\cal
  C} \gamma_5$ and requires a fine tuning of $m \rightarrow m_c$ towards the
continuum limit in order to restore the symmetry. A pair $\xi_1, \xi_2$ of
Majorana fermions may be considered as one Dirac fermion using the
identification $\psi = \frac{1}{\sqrt{2}}(\xi_1 + i \xi_2), \, \bar \psi =
\frac{1}{\sqrt{2}}(\xi_1^T - i \xi_2^T) {\cal C}$ and the corresponding GN
model with $N$ Dirac fermions has an $O(2N)$ flavour symmetry.  At $g=0$,
integrating out the Grassmann variables yields the partition function in terms
of the Pfaffian
\be
Z_\textrm{GN} = \text{Pf}\left[{\cal C} (\gamma_\mu \tilde \partial_\mu +m
 - \frac{1}{2} \partial^* \partial)\right]^{2N}.
\ee
For $g\neq0$ one usually performs a Hubbard-Stratonovich transformation and
introduces a scalar field $\sigma$ conjugate to $\xi^T {\cal C} \xi$ which
yields the new fermionic action
\begin{multline}
 S = \frac{1}{2} \sum_x \xi^T(x) {\cal C} (2 + m + \sigma(x)) \xi(x) \\
- \sum_{x,\mu}  \xi^T(x) {\cal C} \frac{1-\gamma_\mu}{2} \xi(x+\hat\mu),
\end{multline}
together with an additional Gaussian Boltzmann factor $\exp\{- 1/(2 g^2)
\sum_{x} \sigma(x)^2 \}$ for the scalar field.

In order to reformulate the model in terms of closed loops (or equivalently
dimers and monomers) we follow the recent derivation of Wolff
\cite{Wolff:2007ip} (see \cite{Scharnhorst:1996gj,Gattringer:1998cd} for
alternative, but more complicated derivations). One simply expands the
Boltzmann factor for the fermionic fields and makes use of the nil-potency of
the Grassmann variables upon integration.
Introducing $\varphi(x) = 2 + m +\sigma(x)$ and the projectors $P(\pm\mu) = (1
\mp \gamma_\mu)/2$ we can write the fermionic part of the GN path integral (up
to an overall sign) as
\begin{multline}\label{eq:monomer dimer path integral}
\int {\cal D}\xi \, \prod_x  \left( \varphi(x) \xi^T(x) {\cal C}  \xi(x)
\right)^{m(x)} \\
\prod_{x,\mu}\left(\xi^T(x) {\cal C} P(\mu) \xi(x+\hat \mu) 
\right)^{b_\mu(x)}
\end{multline}
where $m(x)=0,1$ and $b_\mu(x)=0,1$ are the monomer and bond (or dimer)
occupation numbers, respectively. Integration over the fermion fields yields
the constraint that at each site $m(x) + \frac{1}{2} \sum_\mu b_\mu(x) = 1$.
Here the sum runs over positive and negative directions and $b_{-\mu}(x) =
b_\mu(x-\hat\mu)$.  The constraint ensures that only closed and
non-intersecting loops of occupied bonds contribute to the partition function
and also accounts for the fact that the loops are non-backtracking, a
consequence of the orthogonal projectors $P(\pm \mu)$.  The weight
$\omega(\ell)$ of each loop $\ell$ can be calculated analytically
\cite{Stamatescu:1980br} and yields $|\omega(\ell)| = 2^{-n_c/2}$ where $n_c$
is the number of corners along the loop. The sign of $\omega$ will generically
depend on the geometrical shape of the loop \cite{Stamatescu:1980br}
prohibiting a straightforward probabilistic interpretation of the loop weights
in dimensions $d>2$.

In two dimensions, however, the sign of the loop only depends on the topology
of the loop, as recently clarified by Wolff \cite{Wolff:2007ip}, and is
determined by the fermionic boundary conditions.  It is therefore useful to
classify all loop configurations into the four equivalence classes ${\cal
  L}_{00},{\cal L}_{10},{\cal L}_{01},{\cal L}_{11}$ where the index denotes
the total winding (modulo two) of the loops in the two directions. The weights
of all configurations in ${\cal L}_{10}$ and ${\cal L}_{11}$ for example will
pick up an overall minus sign if we change the fermionic boundary condition in
the first direction from periodic to antiperiodic, while the weights of the
configurations in ${\cal L}_{00}$ and ${\cal L}_{01}$ remain unaffected. As a
consequence, if we sum over all the topological equivalence classes with
positive weights, i.e., $Z \equiv Z_{{\cal L}_{00}} + Z_{{\cal L}_{10}} + Z_{{\cal
    L}_{01}} + Z_{{\cal L}_{11}}$ we effectively describe a system with
unspecified fermionic boundary conditions. Vice versa, the partition function
$Z_\xi^{10} \equiv Z_{{\cal L}_{00}} + Z_{{\cal L}_{10}} - Z_{{\cal L}_{01}} +
Z_{{\cal L}_{11}}$, e.g., describes a system with fermionic b.c.~antiperiodic
in the first and periodic in the second direction.

Before describing our method to simulate the loop formulation of the GN model,
essentially generating closed loop configurations according to their loop
weight, it is useful to point out the equivalence to the 8-vertex model
\cite{Sutherland:1970,Fan:1970gk}.  The model is formulated in terms of the
eight vertex configurations shown in the top row of Fig.~\ref{fig:8vertex
  tiles} with weights $\omega_i, i=1,\ldots8$.
\begin{figure}[t]
\begin{tabular}{cccccccc}
\includegraphics[width=0.9cm]{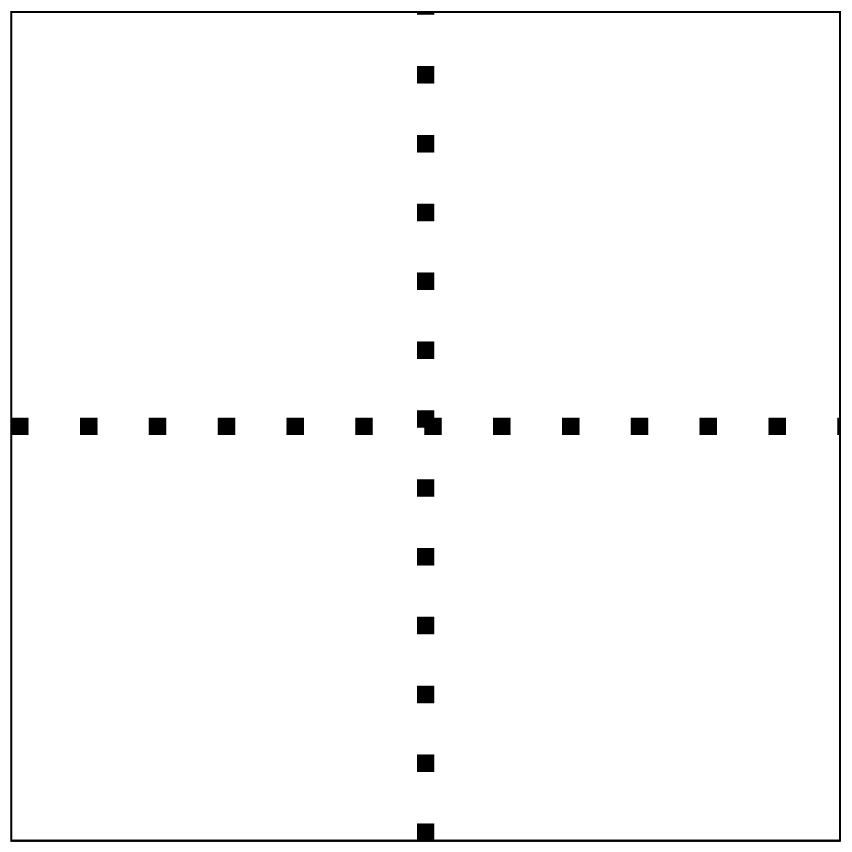} & 
\includegraphics[width=0.9cm]{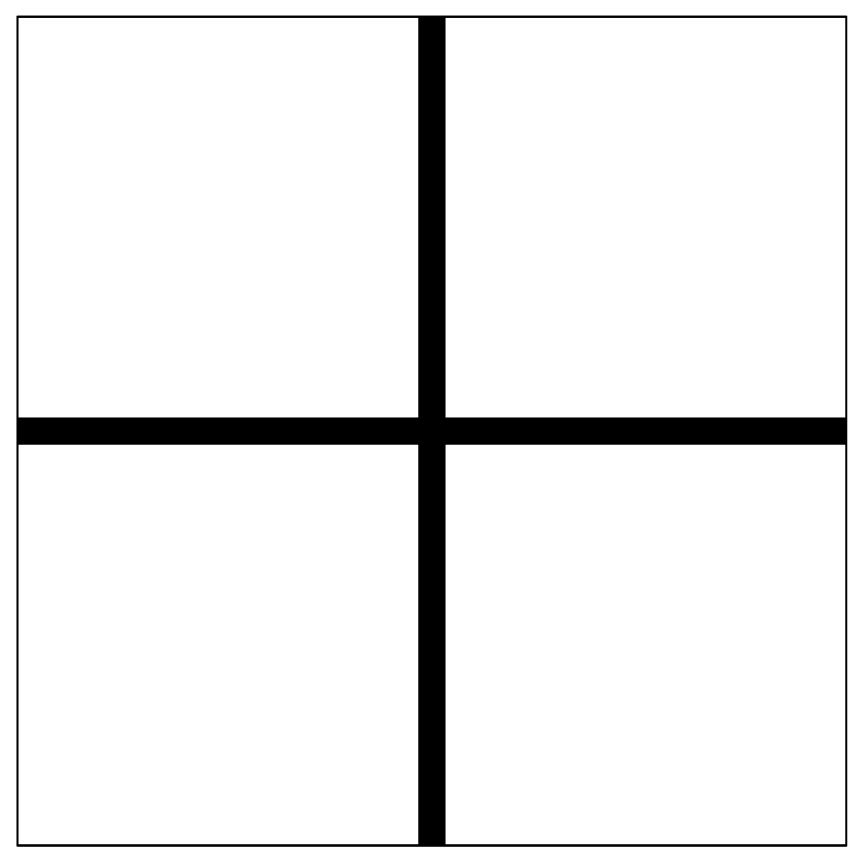} &
\includegraphics[width=0.9cm]{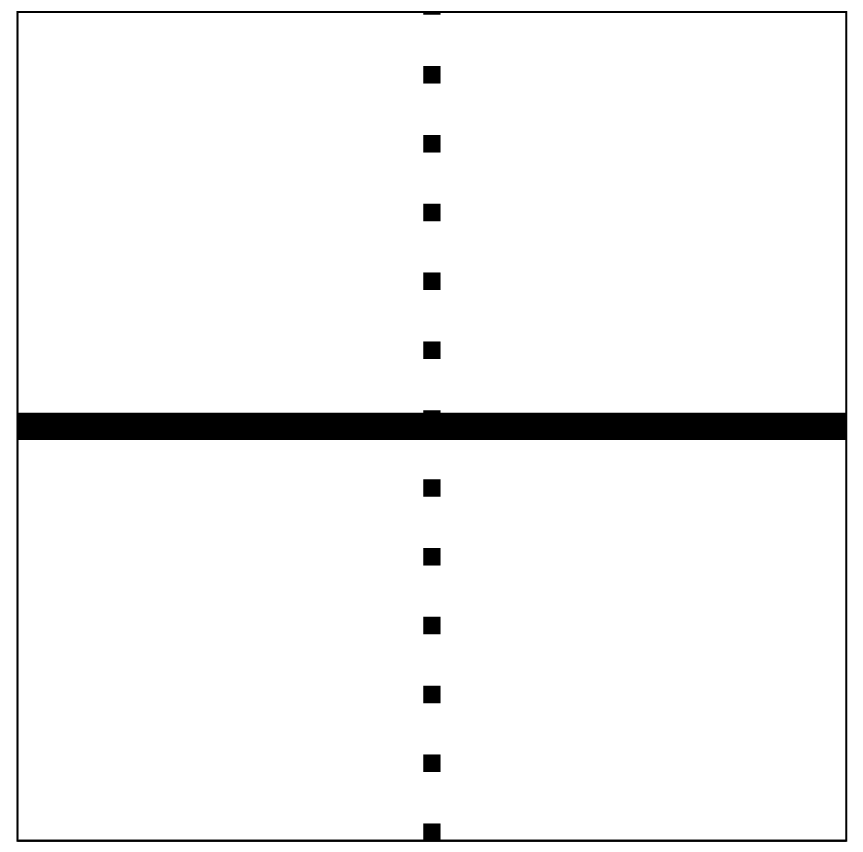} & 
\includegraphics[width=0.9cm]{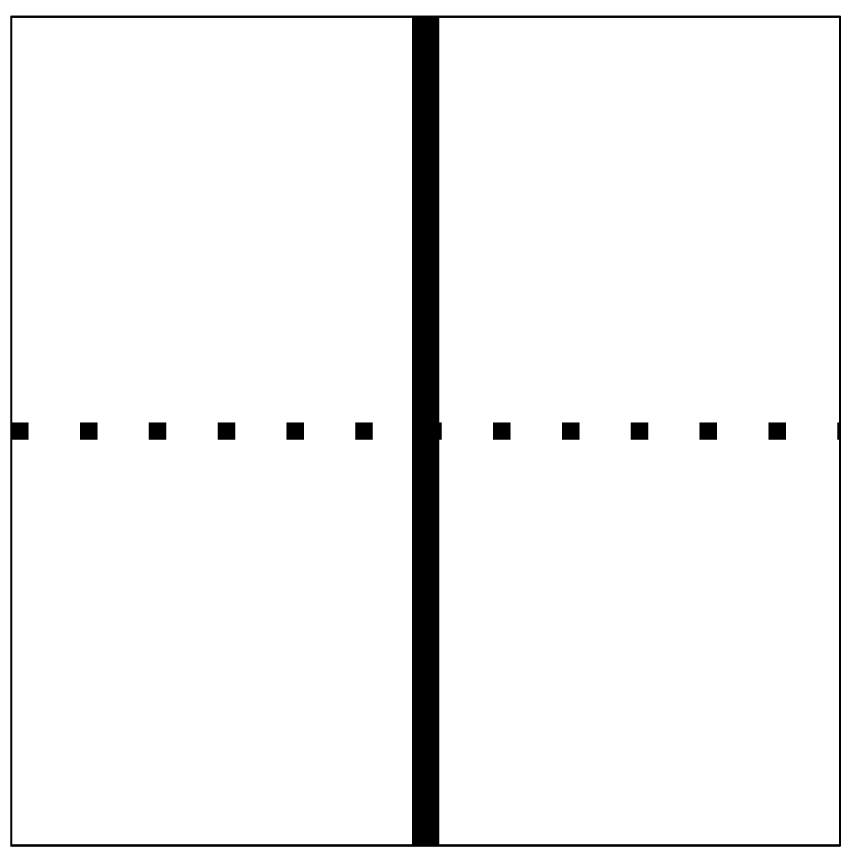} &
\includegraphics[width=0.9cm]{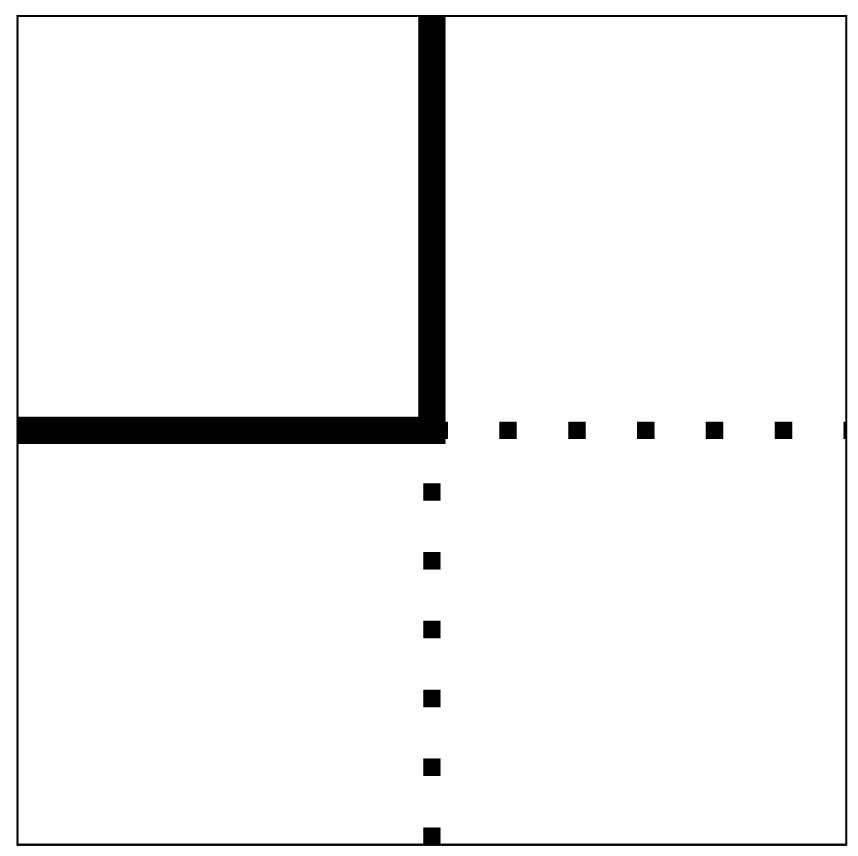} & 
\includegraphics[width=0.9cm]{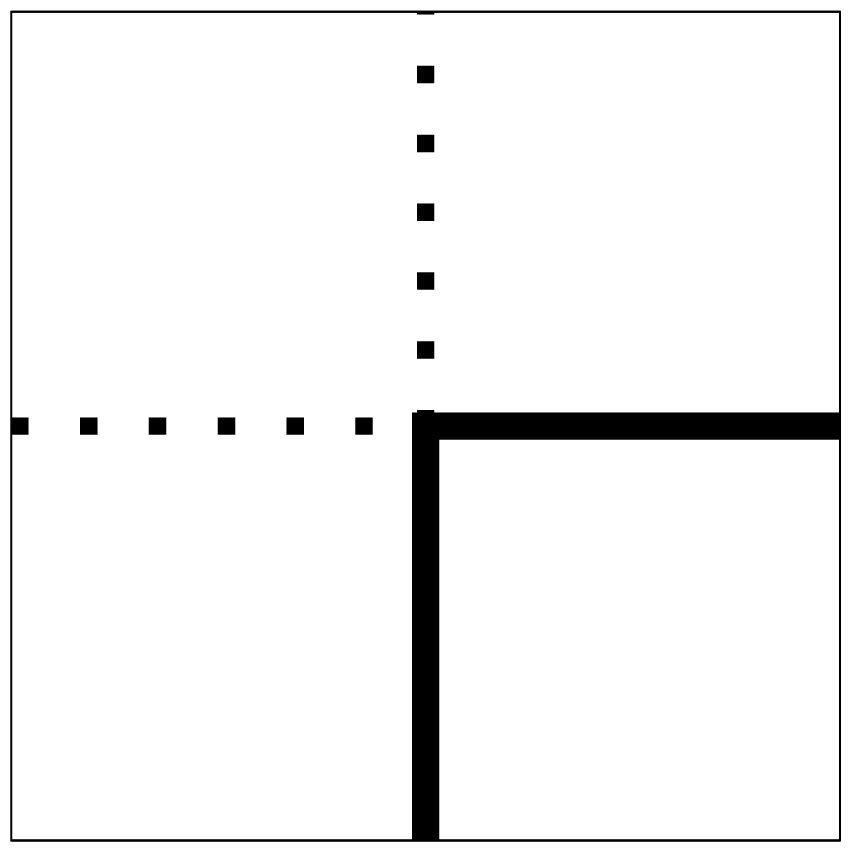} &
\includegraphics[width=0.9cm]{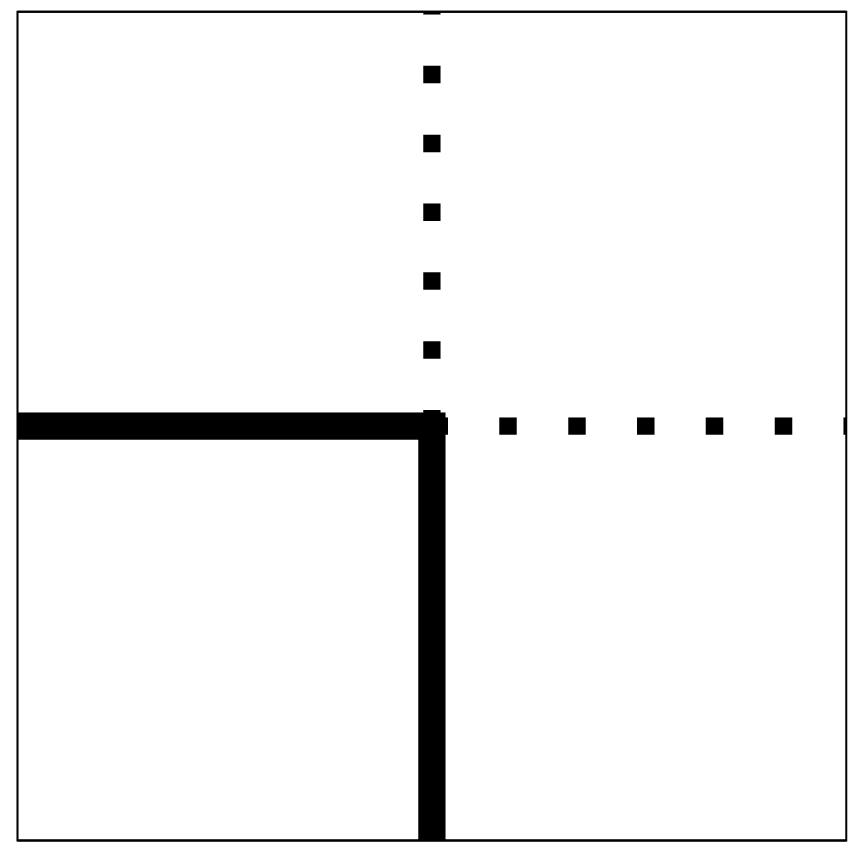} & 
\includegraphics[width=0.9cm]{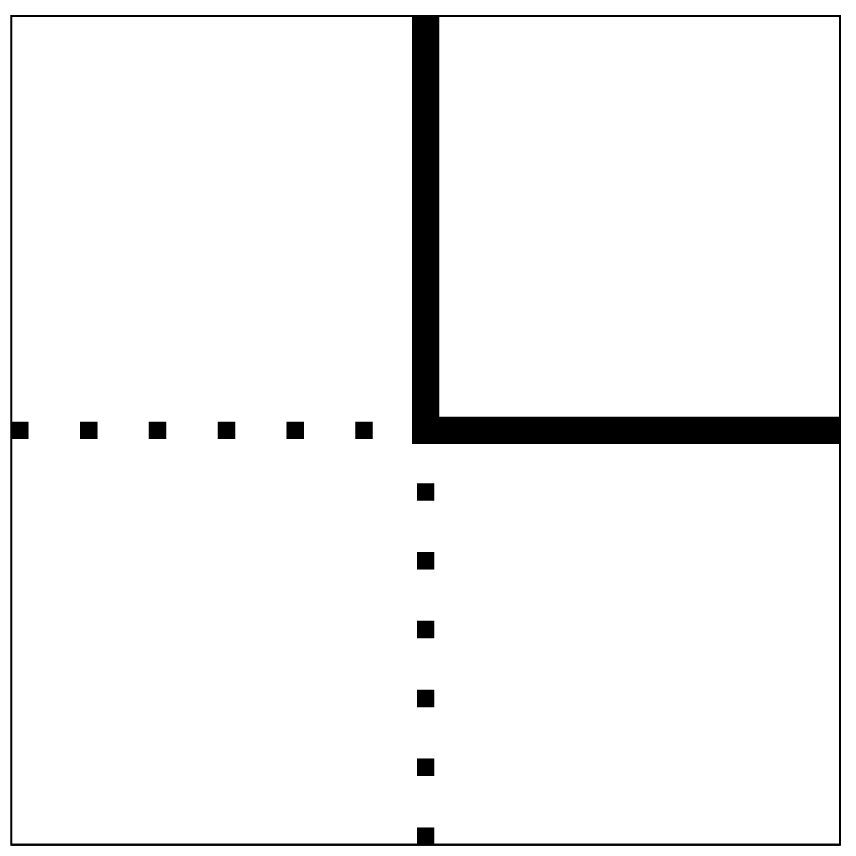} \\
$\omega_1$ & $\omega_2$ & $\omega_3$ & $\omega_4$ & $\omega_5$ & $\omega_6$ &
$\omega_7$ & $\omega_8$ \\
&&&&&&& \\
\includegraphics[width=0.9cm]{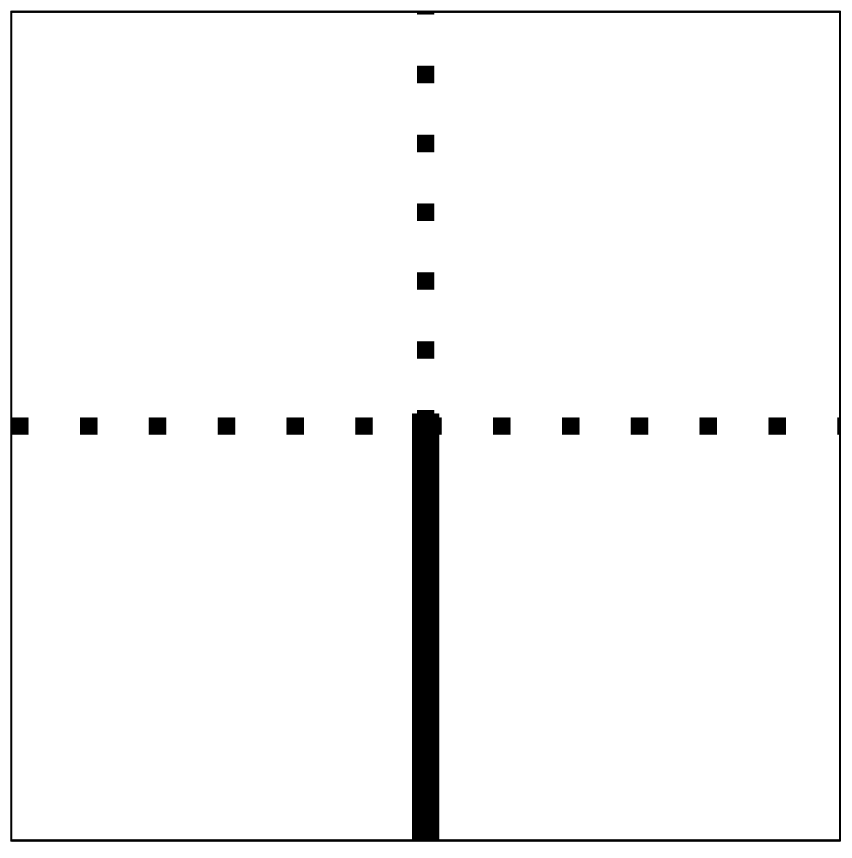} & 
\includegraphics[width=0.9cm,angle=90]{Figures/vertex_head} &
\includegraphics[width=0.9cm,angle=180,origin=c]{Figures/vertex_head} & 
\includegraphics[width=0.9cm,angle=-90,origin=c]{Figures/vertex_head} &
\includegraphics[width=0.9cm]{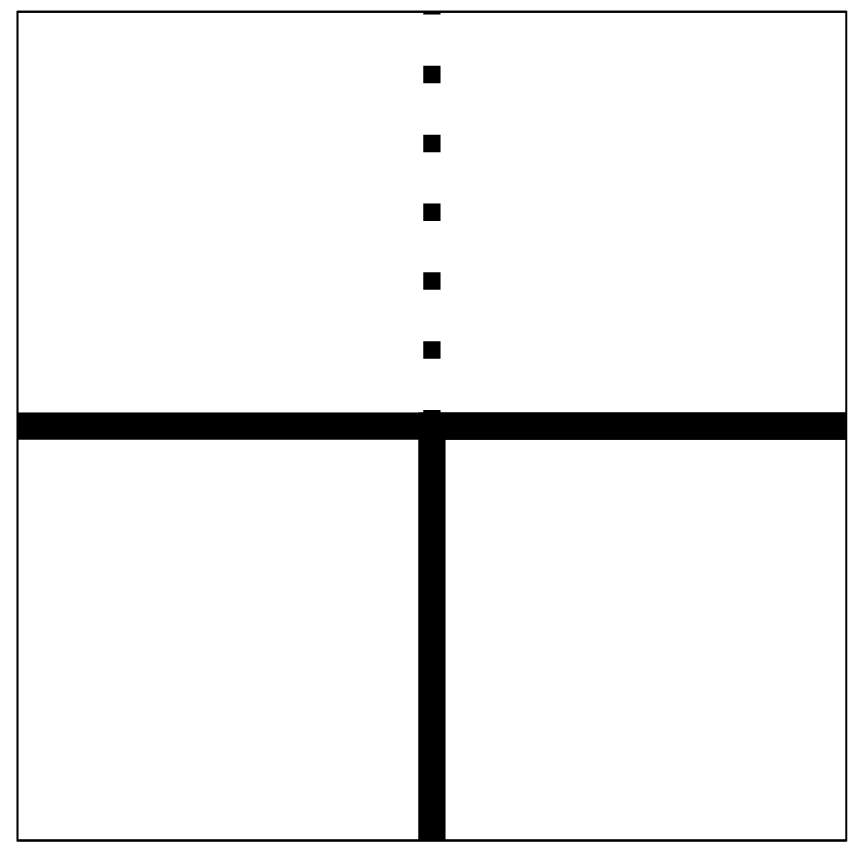} & 
\includegraphics[width=0.9cm,angle=90]{Figures/vertex_reconnect} &
\includegraphics[width=0.9cm,angle=180,origin=c]{Figures/vertex_reconnect} & 
\includegraphics[width=0.9cm,angle=-90,origin=c]{Figures/vertex_reconnect} \\
$\omega_9$ & $\omega_{10}$ & $\omega_{11}$ & $\omega_{12}$ & $\omega_{13}$ & 
$\omega_{14}$ & $\omega_{15}$ & $\omega_{16}$ 
\end{tabular}
\caption{ 
The vertex configurations and weights of the eight-vertex model (top
  row) and the extended model (bottom row).  
}
\label{fig:8vertex tiles}
\end  {figure}
The partition function is then defined as the sum over all possible tilings of
the square lattice with the eight vertices such that only closed (but possibly
intersecting) paths occur. To be precise, one has
\be
Z_\textrm{8-vertex} = \sum_\textrm{CP} \prod_{x} \omega_i(x). 
\ee
where the sum is over all closed path configurations ($\textrm{CP}$) and the
weight of each configuration is given by the product of all vertex weights in
the configuration. For the GN model we have the following weights
\be
\begin{tabular}{rclcrcl}
$\omega_1$ &$=$& $\varphi(x)$,& $\quad$ &  $\omega_3=\omega_4$ &$=$& 1, \\
 $\omega_2$ &$=$& 0,& $\quad$ & 
 $\omega_5=\omega_6=\omega_7=\omega_8$ &$=$& $\frac{1}{\sqrt{2}}$,
\end{tabular}
\ee
i.e.~each corner contributes a factor $1/\sqrt{2}$, while crossings of two
lines are forbidden ($\omega_2=0$) and each empty site carries the monomer
weight $\omega_1 = \varphi(x)$. From here it also becomes clear that for a
single Majorana fermion the interaction term proportional to $g$ is
irrelevant. Since the partition function is now factorised into terms at each
$x$ we can integrate over the scalar field $\sigma(x)$ at each site
separately. However, since the integration measure is even in $\sigma(x)$ the
term linear in $\sigma(x)$ will not survive and the monomer weight reduces to
$\omega_1 = 2 + m$. This is how the free Majorana fermion is recovered after
the Hubbard-Stratonovich transformation. If one considers two or more Majorana
flavours which are coupled through the four-fermion interaction, the
integration over $\sigma(x)$ becomes non-trivial, but can still be done
analytically \cite{Gattringer:1998cd} yielding 8-vertex models coupled to each
other. It should be stressed, however, that from an algorithmic point of view
the generic case with an arbitrarily varying field $\sigma(x)$ is equally
accessible and involves no complication whatsoever.

The fact that the GN model with a single Majorana fermion is effectively a
free fermion system expresses itself also through the vertex weights
fulfilling the free fermion condition \cite{Fan:1969wb,Fan:1970gk} $\omega_1
\omega_2 + \omega_3 \omega_4 = \omega_5 \omega_6 + \omega_7 \omega_8$.
8--vertex models fulfilling the free fermion condition as well as those in
zero field \cite{Baxter:1971cr,Baxter:1982} are analytically solvable, in
particular also the standard Ising model. We use this fact to our convenience
and use the Majorana GN model as a benchmark which allows to compare the
results of our algorithm with analytic results.  Instances of the $8$--vertex
model for which no analytic solutions are known, but are accessible with our
algorithm, include the Ising model with additional next-to-nearest-neighbour
and quartic interactions \cite{Baxter:1982}. Another $8$--vertex model with a
fermionic interpretation is the one-flavour Schwinger model with Wilson
fermions in the strong coupling limit
\cite{Salmhofer:1991cc,Gattringer:1999hr}. The vertex weights are given by
\be
\begin{tabular}{rclcrcl}
$\omega_1$ &$=$& $(m+2)^2$,& $\quad$ &  $\omega_3=\omega_4$ &$=$& 1, \\
 $\omega_2$ &$=$& 0,& $\quad$ & 
 $\omega_5=\omega_6=\omega_7=\omega_8$ &$=$& $\frac{1}{2}$,
\end{tabular}
\ee
where the monomer weight and the corner weights are squared due to the fact
that we are dealing with a pair of Majorana fermions glued together
\footnote{One can in fact derive these weights also for $d>2$ and show that
  all loop contributions are positive, hence allowing simulations of the
  Schwinger model in the strong coupling limit in arbitrary dimensions.}.

Let us now turn to the description of the new method to efficiently simulate
any vertex model in arbitrary dimensions with generic (positive) weights
$\omega_i$, including the fermionic models discussed above.  For illustrative
purpose we restrict the discussion to the 8--vertex model.  The method is an
extension of the so-called worm algorithm by Prokof'ev and Svistunov
\cite{Prokof'ev:2001zz}. The configuration space of closed loops is enlarged
to contain also open strings.  For the GN model such an open string with ends
at $x$ and $y$ corresponds to the insertion of two Majorana fields at
positions $x$ and $y$ which is simply the Majorana fermion propagator
\be \label{eq:open string} 
G(x,y) = \int {\cal D} \xi \, 
      e^{-S_\textrm{GN}} \,\, \xi(x) \xi(y)^T {\cal C} \, .  
\ee 
Similar interpretations of the open string can be obtained for other vertex
models.  The open string is now the basis for a Monte Carlo algorithm which
samples directly the space of 2-point correlation functions instead of the
standard configuration space.  This is the reason why the algorithm is capable
of beating critical slowing down as demonstrated below: at a critical point
where the correlation length grows large, the configurations are updated
equally well on all length scales up to a scale of the order of the
correlation length.

In the vertex language the insertions correspond to the new vertex
configurations depicted in the bottom row of Fig.~\ref{fig:8vertex tiles}. A
configuration containing a single open string corresponds to a loop
configuration with two instances of vertex 9-16 which are connected by a
string. Note that vertices 13-16, while present in the generic extended
vertex-model, do not have a physical interpretation in terms of fermionic
fields since they are explicitely forbidden by Pauli's exclusion principle
(fermionic lines are not allowed to intersect).  Nevertheless they can also be
included in the fermionic models, simply for algorithmic efficiency, and we do
so in our implementation.

The algorithm now proceeds by locally updating the ends of the open string
using a simple Metropolis or heat bath step according to the weights of the
corresponding 2-point function. When one end is shifted from, say, $x$ to one
of its neighbouring points $y$, a dimer on the corresponding bond is destroyed
or created depending on whether the bond is occupied or not. In the process,
the two vertices at $x$ and $y$ are changed from $v_x, v_y$ to $v'_x, v'_y$
and the move is accepted with probability
\be 
P(x\rightarrow y) = \min\left[1,\frac{\omega_{v'_x}}{\omega_{v_x}} 
\frac{\omega_{v'_y}}{\omega_{v_y}}\right] 
\ee
in order to satisfy detailed balance. So a global update results from a
sequence of local moves, and in this sense it is similar in spirit to the loop
cluster update suggested in \cite{Evertz:1992rb}.

Whenever the two ends of the open string meet, a new closed loop is formed and
the new configuration contributes to the original partition function $Z$ in
one of the classes ${\cal L}_{00},{\cal L}_{10},{\cal L}_{01},{\cal L}_{11}$.
In this way the overall normalisation is ensured, and expectation values can
be calculated as usual. From here it also becomes clear that the algorithm
switches between the topological sectors with ease: as the string evolves it
can grow or shrink in any direction and wrap around the torus as many times as
it likes. Effectively, the algorithm simulates a system with fluctuating
topological boundary conditions.
\begin{figure}[t]
\includegraphics[width=\linewidth]{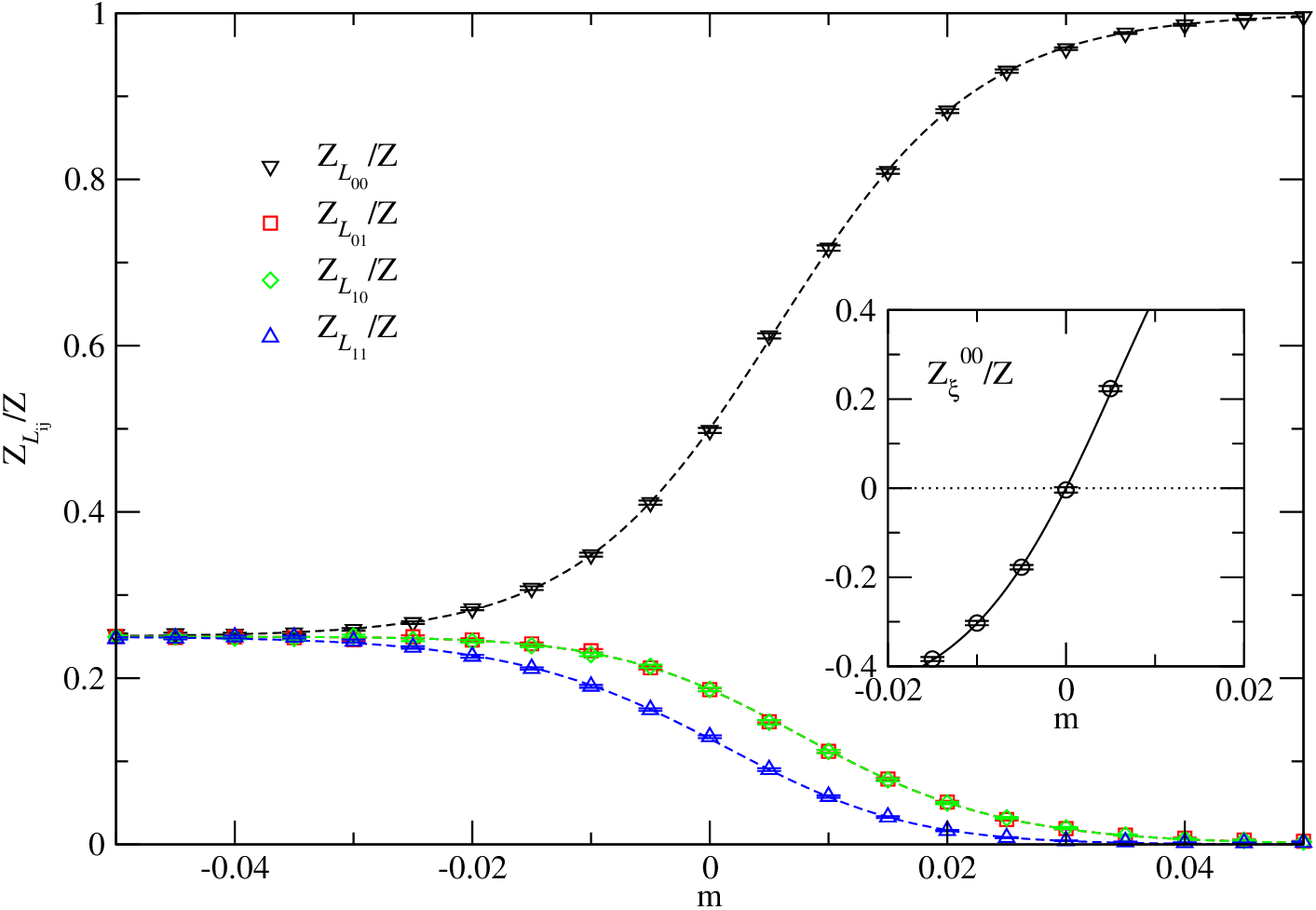}
\caption{Results for the ratios $Z_{{\cal L}_{ij}}/Z$ with $i,j=0,1$ for the
  Majorana GN model on a $128^2$ lattice as a function of the bare mass $m$.
  Dashed lines are the exact results calculated from the Pfaffians. Note that
  the curves and the data for $Z_{{\cal L}_{10}}/Z$ and $Z_{{\cal L}_{01}}/Z$
  lie on top of each other. The inset shows the partition function ratio
  $Z_\xi^{00}/Z$ and illustrates how the zero mode at $m=0$ is reproduced.}
\label{fig: GN partition function ratios}
\end{figure}

In principle, the weight of the open string can be chosen arbitrarily, but the
physical interpretation given by eq.(\ref{eq:open string}) suggests to choose
the weights $\omega_9$ to $\omega_{16}$ such that the open string
configurations sample directly the 2-point correlation function. The open
string configurations then provide an improved estimator for the 2-point
correlation function \footnote{Note that one can also sample $(2n)$-point
  correlation functions.}. During the simulation one simply updates a table
for $G(x,y)$ as the string endpoints move around and the expectation value is
obtained by forming $\langle G(x,y) \rangle_Z = G(x,y)/Z$. For the fermionic
models we also need to keep track of the Dirac structure associated with
$G(x,y)$. This is most easily done by adding the product of the Dirac
projectors along the string $\ell$, i.e.~$\prod_{\mu \in {\ell}} P(\mu)$, as a
contribution at each step. Care has to be taken when the open string winds an
odd times around a boundary on which we want to impose antiperiodic boundary
conditions for the fermions. In that case we need to account for an additional
minus sign in the contribution to $G(x,y)$.
For the fermionic models where vertices 13-16 have no physical meaning, the
weights $\omega_{13}$ to $\omega_{16}$ can be tuned for algorithmic efficiency
and do not follow any physically inspired rule. A good choice is to use the
geometric mean of the weights $\omega_i$ of those vertices that can be reached
in one further update step, e.g.~$\omega_{13} = ( \omega_4 \omega_6
\omega_7)^{1/3}$.  Finally, let us emphasise again that the algorithm
described here is applicable to any vertex model, also in higher dimensions,
as long as the weights are positive definite in well defined configuration
classes.

\begin{figure}[t]
\includegraphics[width=\linewidth]{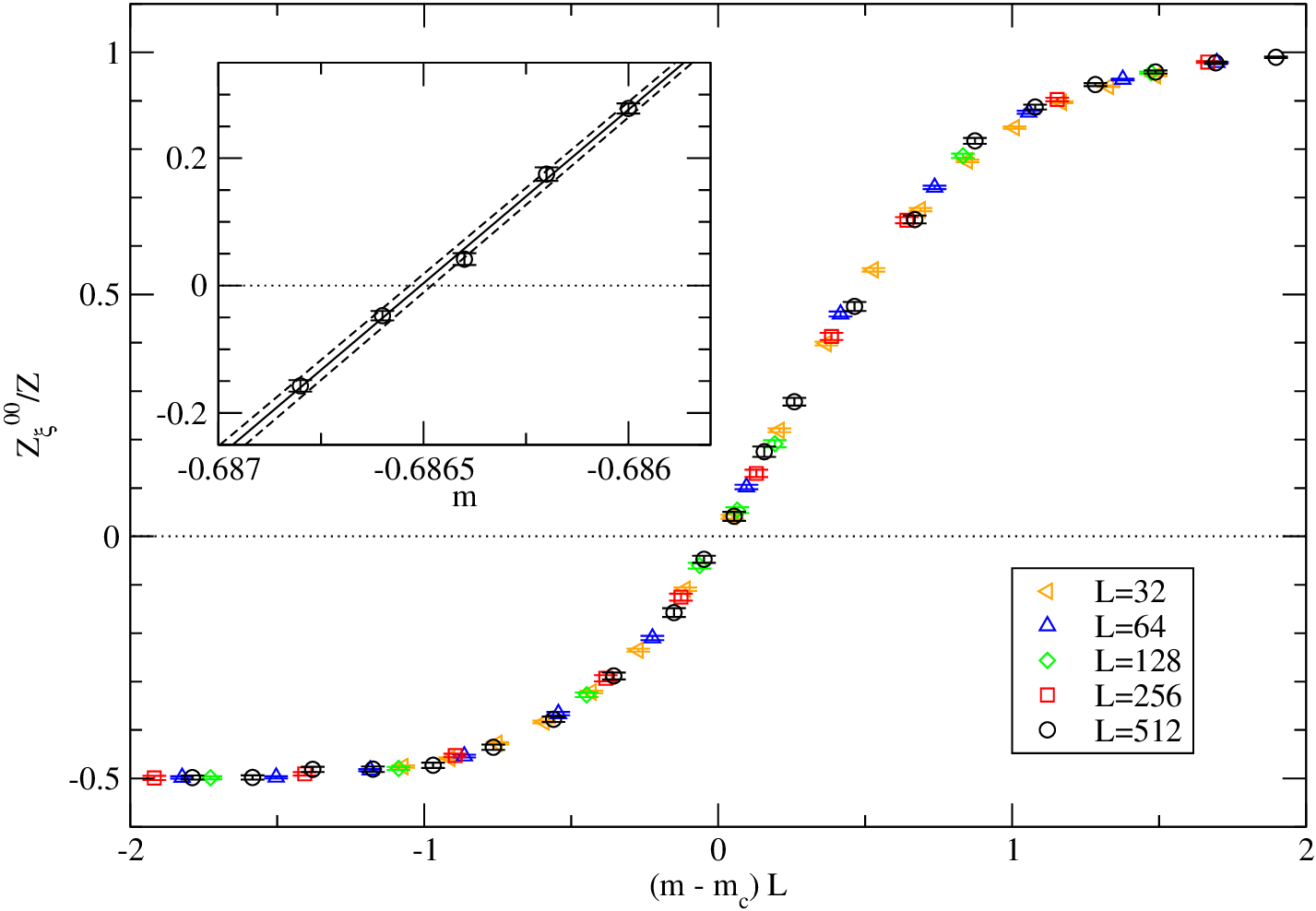}
\caption{Results for the ratio $Z_\xi^{00}/Z$ in the Schwinger model at
  strong coupling for various system sizes. The inset shows the determination
  of the critical mass $m_c = -0.686506(27)$ while the main plot shows
  the collapse from finite size scaling consistent with a second order phase
  transition in the universality class of the Ising model (with critical
exponent $\nu=1$).}
\label{fig:scsm_Zpp}
\end{figure}
We have performed extensive tests of our algorithm by comparing to exact
results known from Pfaffians (for the Majorana GN model) or from explicit
calculations on small lattices. Simple observables are linear combinations of
partition functions and ratios thereof, e.g.~$Z_{{\cal L}_{ij}}/Z$ with
$i,j=0,1$. In Fig.~\ref{fig: GN partition function ratios} we show the results
for the ratios $Z_{{\cal L}_{ij}}/Z$ in the Majorana GN model on a $128^2$
lattice as a function of the bare mass $m$. Dashed lines are the exact results
calculated from the Pfaffians. Note that all partition function ratios are
obtained in the same simulation. In the inset we also show the ratio
$Z_\xi^{00}/Z$ where $Z_\xi^{00} \equiv Z_{{\cal L}_{00}} - Z_{{\cal L}_{01}} -
Z_{{\cal L}_{10}} - Z_{{\cal L}_{11} }$ is the partition function with
fermionic b.c.~periodic in space and time direction. In that situation the
Majorana Dirac operator has a zero mode at $m=0$ (and at $m=-2$) and the
system is critical.  The inset in Fig.~\ref{fig: GN partition function ratios}
illustrates that the algorithm can reproduce this zero mode without problems
and that we can in fact simulate directly at the critical point.  Conversely,
we can use $Z_\xi^{00}/Z = 0$ as a definition of the critical point $m=m_c$.
In Fig.~\ref{fig:scsm_Zpp} we show our results for $Z_\xi^{00}/Z$ as a
function of the bare mass $m$ in the Schwinger model in the strong coupling
limit for various volumes. The critical point can be determined accurately
with very little computational effort and we obtain $m_c=-0.686506(27)$
(cf.~inset in Fig.~\ref{fig:scsm_Zpp}) from our simulations on the largest
lattice with $L=512$. Further improvement could be achieved by employing
standard reweighting techniques as done in \cite{Gausterer:1995jh} where they
obtained $m_c=-0.6859(4)$. These calculations indicated a second order phase
transition in the universality class of the Ising model (with critical
exponent $\nu\simeq 1$). Our results in Fig.~\ref{fig:scsm_Zpp} now confirm
this by demonstrating that the partition function ratios $Z_\xi^{00}/Z$ as a
function of the rescaled mass $(m - m_c) L^\nu$ with $\nu=1$ beautifully
collapse onto a universal scaling curve.

\begin{figure}[t]
\includegraphics[width=\linewidth]{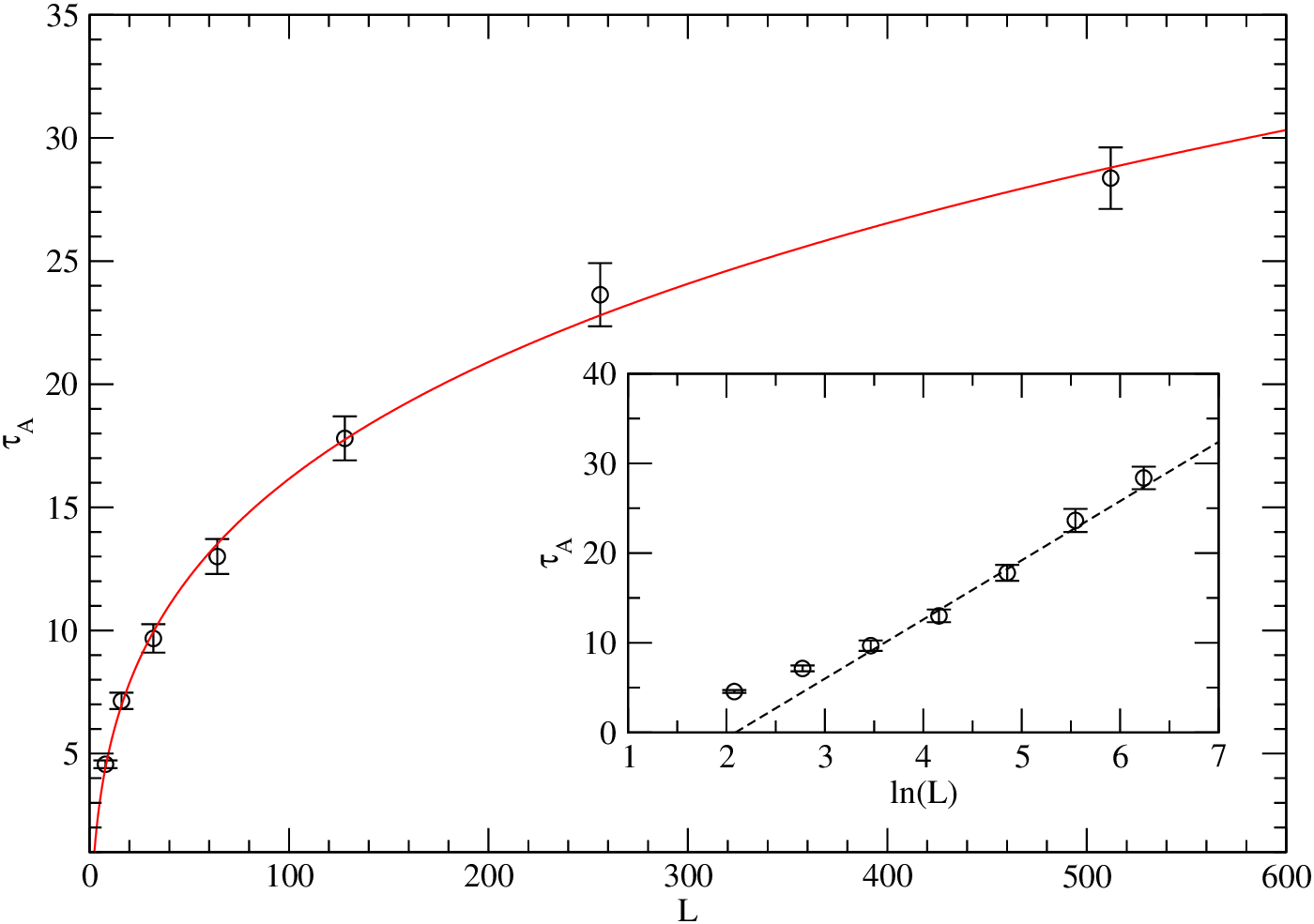}
\caption{Integrated autocorrelation time $\tau_A$ of the energy for the
  Schwinger model in the strong coupling limit as a function of the system
  size $L$ at the critical point $m=m_c$. The line is a fit $\tau_A \propto
  L^z$ yielding $z=0.25(2)$. The inset shows the logarithmic dependence
  $-18(3) + 7.5(6) \ln(L)$ from fitting to $L \geq 64$.}
\label{fig:autocorrelation_time}
\end{figure}
The efficiency of the algorithm and the fact that critical slowing down is
essentially absent is demonstrated in Fig.~\ref{fig:autocorrelation_time}
where we show the integrated autocorrelation time $\tau_A$ of the energy as a
function of the linear system size $L$ at the critical point $m=m_c$. (Similar
plots can be obtained for the Majorana GN model.)  The functional dependence
on $L$ can be well fitted ($\chi^2/\textrm{dof}=1.28)$ by $\tau_A \propto L^z$
all the way down to our smallest system size $L=8$. We obtain $z=0.25(2)$
which is consistent with just using the largest two system sizes. It is an
amazing result that our local Metropolis-type update appears to have a
dynamical critical exponent close to zero.  The autocorrelation time may also
depend logarithmically on $L$ and a fit to $L \geq 64$ yields $-13.8(1.9) +
6.6(4) \ln(L)$ with $\chi^2/\textrm{dof}=1.00$.

In conclusion, we have presented a new type of algorithm for generic vertex
models. It relies on sampling directly 2-point correlation functions and
essentially eliminates critical slowing down. We have successfully tested our
algorithm on the Majorana GN model and on the Schwinger model in the strong
coupling limit and found remarkably small dynamical critical exponents.  The
algorithm definitely opens the way to simulate efficiently generic vertex
models (with positive weights) in arbitrary dimensions, in particular the GN
model with any number of flavours, the Thirring model, the Schwinger model in
the strong coupling limit (in arbitrary dimensions), as well as fermionic
models with Yukawa-type scalar interactions, all with Wilson fermions.

\acknowledgments

I would like to thank Philippe de Forcrand and Michael Fromm for useful and
sometimes crucial discussions. This work is supported by SNF grant
PP002-\_119015.

\bibliography{esrfvm}

\end{document}